\documentclass[review,12pt]{elsarticle}

\usepackage{algorithmic}
\usepackage{graphicx}
\usepackage{textcomp}
\usepackage{xcolor}

\makeatletter
\def\ps@pprintTitle{%
  \let\@oddhead\@empty
  \let\@evenhead\@empty
  \let\@oddfoot\@empty
  \let\@evenfoot\@oddfoot
}
\makeatother

\usepackage{array}
\usepackage[caption=false,font=normalsize,labelfont=sf,textfont=sf]{subfig}
\usepackage{stfloats}
\usepackage{url}
\usepackage{verbatim}
\usepackage{balance}
\usepackage{xspace}
\usepackage{booktabs}
\usepackage{multirow}
\usepackage{enumitem}
\usepackage{url}
\usepackage{comment}

\begin{document}

\begin{frontmatter}

\title{Detection of Compromised Functions in a Serverless Cloud Environment
}

\author{Danielle Lavi, Oleg Brodt, Dudu Mimran, Yuval Elovici, Asaf Shabtai}

\let\theaffn=\relax
 \affiliation{organization={Dept. of Software and Information Systems Engineering \\ Ben-Gurion University of the Negev} }

\begin{abstract}

Serverless computing is an emerging cloud paradigm with serverless functions at its core.
While serverless environments enable software developers to focus on developing applications without the need to actively manage the underlying runtime infrastructure, they open the door to a wide variety of security threats that can be challenging to mitigate with existing methods.
Existing security solutions do not apply to all serverless architectures, since they require significant modifications to the serverless infrastructure or rely on third-party services for the collection of more detailed data. 
In this paper, we present an extendable serverless security threat detection model that leverages cloud providers' \emph{native monitoring tools} to detect anomalous behavior in serverless applications.
Our model aims to detect compromised serverless functions by identifying post-exploitation abnormal behavior related to different types of attacks on serverless functions, and therefore, it is a last line of defense. 
Our approach is not tied to any specific serverless application, is agnostic to the type of threats, and is adaptable through model adjustments.
To evaluate our model's performance, we developed a serverless cybersecurity testbed in an AWS cloud environment, which includes two different serverless applications and simulates a variety of attack scenarios that cover the main security threats faced by serverless functions.
Our evaluation demonstrates our model's ability to detect all implemented attacks while maintaining a negligible false alarm rate.

\end{abstract}

\end{frontmatter}

\section{\label{sec:intro}Introduction}
Serverless computing is an increasingly popular cloud computing paradigm that enables organizations to build and deploy software and services in the cloud, without the need to maintain, provide, or scale resources like physical or virtual servers~\cite{serverlessSurvey}, and to pay only for the resources their applications consume.
It primarily provides cloud usage models centered around Backend-as-a-Service (BaaS) and Function-as-a-Service (FaaS)~\cite{serverlessSurveyNupponen}, with Function-as-a-Service being the predominant serverless model~\cite{hassan2021survey}.
Within the context of Function-as-a-Service (FaaS), serverless functions (e.g., AWS Lambda~\cite{AWS_Lambda_Documentation}, Microsoft Azure Functions~\cite{Microsoft_Azure_Functions}, Google Cloud Functions~\cite{Google_Cloud_Functions}) are compact, stateless, single-purpose components of business logic within a serverless application, and are intended to be activated in response to predefined events.

The rapid shift from on-premises to cloud computing and serverless computing has presented cybersecurity defenders with a wider attack surface~\cite{owasp_top10} and a growing number of challenges~\cite{zissis2012addressing}, some of which are described below.

\noindent\textbf{Limited access to the underlying infrastructure.} 
In serverless computing, organizations typically operate within predefined infrastructure configurations, lacking the capability to customize them to meet their specific security requirements.
As a result, security teams cannot add lower-level defenses or monitoring capabilities, and they must rely on the existing measures and logging provided by the cloud platform.
Thus, their ability to assess their system's threats and determine the organization's defense posture is limited~\cite{li2022securing,ibrahim2016cloudInfra}.

\noindent\textbf{Short development cycles.}
In an agile application development setting, many code changes are made, and new resources are often created, making the application more dynamic, which in turn makes application monitoring much more challenging.
As serverless applications' code changes frequently, maintaining a consistent security posture is challenging due to a constantly shifting attack surface and changes in logging data, application configurations, and permissions.

\noindent\textbf{Large and complex attack surface.}
While serverless applications have inherited the security problems typically associated with cloud computing (e.g., configuration mismanagement, identity management, data leakage), they also have their own security issues, resulting in a larger attack surface~\cite{pusuluri2022taxonomy,marin2022serverless}. 
In most cases, serverless functions perform a single atomic task and are able to run in parallel with other functions. 
Therefore, thousands of serverless functions may use the same resources at the same time and need to communicate with one another and with additional services, which creates a complex and wide attack surface. 
In addition, serverless functions may be invoked by a plethora of internal and external events, providing an attacker with a variety of avenues to try and exploit.

Although the challenges of and threats to serverless computing are well documented, in many cases, the research community has yet to provide effective solutions~\cite{li2022securing}.
The solutions proposed in prior research in the field relied on rule-based methods~\cite{obetz2019static}, required the cooperation of third-party services~\cite{alpernas2018secure, datta2020valve}, or demanded modifications that could not be made in a \emph{fully managed environment}~\cite{datta2020valve,sankaran2020workflow,datta2022alastor,jeganguarding}.
The majority of the suggested approaches involve either static or dynamic analysis of applications. 
This analysis aims to identify irregularities in the applications' data flow and to enforce access control measures.
For example, static analysis solutions can be used to audit and detect ``bad practices" concerning serverless security~\cite{obetz2019static,lambdaGuard}. 
However, this approach does not detect external threats (e.g., injections) or scale well in fast-moving, dynamic environments.
Furthermore, static analysis techniques may misclassify malicious code as benign, as these methods can encounter difficulties in distinguishing subtle distinctions between malicious and legitimate code.

Previous research that employed dynamic analysis techniques used \emph{modified} versions of the functions' runtime environment to monitor and detect malicious activity~\cite{datta2020valve,sankaran2020workflow,datta2022alastor,jeganguarding}.
Those methods, however, do not address the need for threat detection solutions in \emph{a fully managed serverless environment}. 
Their ability to control and modify the existing infrastructure is only possible when operating in a \emph{non-managed environment}, which is unrealistic in the major cloud providers' settings.
Moreover, the approach adopted by many of these solutions~\cite{datta2020valve,sankaran2020workflow,jegan2020guarding,jeganguarding} in countering injection attacks or the introduction of malicious code through the supply chain, involves the blocking of suspicious requests, rather than actively identifying and flagging malicious behavior, and take the ``presumption of compromise" approach.
In addition to the fact that prevention is usually prone to failure, this approach could also lead to a high false alarm rate, which would interfere with the application's availability.

While serverless computing poses many security challenges, it also offers new threat detection opportunities, since atomic units of serverless functions enable native function-level logging with high granularity.
Among other things, high-resolution logging allows application administrators to gain a deeper understanding of serverless applications and security teams to identify anomalies and potential threats. 
In this study, we address threats to serverless functions by leveraging native function-level logging. 
There are several issues arising when developing a threat detection model for serverless functions, including issues stemming from their dynamic nature, high volume, and the wide variety of activities that take place during their lifetimes.
These create a lot of noise in a system and enable attackers to ``stay below the radar."

In this paper, we propose an unsupervised deep learning anomaly detection model that relies solely on the \emph{cloud provider's native monitoring tools} to detect abnormal behavior in serverless applications.
Rather than addressing the many ways in which a system can be attacked, our model concentrates on an attack's impact on the application's behavior.
It is designed to detect abnormal behavior as manifested in the native activity logs while being agnostic to the type of threat (known or unknown).

Our detection model can be used in an online or offline mode. 
Regardless of the mode, the model learns the normal behavior of the serverless environment, specifically, serverless functions. 
In an online mode, the model is used continuously to issue alerts for anomalies in real-time. 
In an offline mode, the model is only used when the enterprise requires hyper-logging and hyper-monitoring capabilities, as in the case of incident response, to allow responders to monitor and quickly zero in on abnormal behavior. 
Since there is often a trade-off between security and resource utilization, each organization can determine the appropriate mode of operation based on its security policy and risk assessment. 

Our model does not require any changes to the existing infrastructure, as it uses the monitoring tools that exist in cloud environments.
The unique features extracted by our threat detection model enrich the basic information provided by the activity logs and create a more in-depth informative perspective on the behavior of \emph{serverless functions}.

To evaluate the proposed model, we developed a serverless evaluation testbed in an AWS environment that contains two open-source serverless applications.
As part of the testbed, we simulated benign user activity as well as a variety of attacks. 
In our evaluation, the model successfully detected all of the implemented attacks while maintaining a very low false alarm rate (0.003).
Such capabilities are crucial for improving the security of today's widely used serverless platforms and the performance of the security teams responsible for defending them.

To summarize, the contributions of our paper are as follows:  
\begin{itemize}
    \item We present a novel threat detection model specifically designed for the identification of compromised serverless functions.
    The proposed model is built on top of cloud providers' native hyper-logging and monitoring tools, and security teams can use it without the need for external collaboration (e.g., asking a developer to change the source code or enlisting the cooperation of third-party services).
    \item Our approach is serverless application agnostic, threat agnostic, easy to expand with model fine-tuning, and easy to maintain, since organizations can use the same model to monitor multiple functions and applications (instead of creating a model for each function).
    \item We collected a unique anomaly detection dataset that contains data reflecting the activity of two different serverless applications, each of which was developed in a different serverless paradigm and has its own unique characteristics.
   \item We have made the dataset, testbed, code, and application activity simulator available to the research community (the relevant details will be published with this paper).

\end{itemize}

\section{\label{sec:threat}Threat Model}

In addition to being prone to cloud-related threats, serverless environments have a unique threat landscape of their own~\cite{pusuluri2022taxonomy, marin2022serverless}.  
Serverless functions, which are the main component of serverless environments, provide attackers with new opportunities to exploit the functions' permissions and accomplish their goals~\cite{pusuluri2022taxonomy, marin2022serverless,li2022securing,chawla2021information}.
These functions are often composed of different components and services, creating many potential entry points for attackers (e.g., the function's code or dependencies, the underlying infrastructure). 
Serverless functions are event-driven and can be invoked in response to a wide variety of events (e.g., HTTP requests, database updates, storage modifications), which increases their attack surface.
They are also stateless and short-lived.
As a result, it is complicated to determine their role within the application flow and the context in which they run.

In addition to the security complexity associated with the intrinsic nature of serverless functions, misconfiguration errors may also play a significant role. 
A misconfigured function (e.g., over-privileged, having an excessively long timeout) may result in vulnerabilities that allow an attacker to exploit the function~\cite{wen2023rise}.
In our research, we define a compromised serverless function as one that acts in a way that was unintended as a result of the action of a malicious actor.
Our challenge is to detect compromised functions among a large number of functions running simultaneously and operating on the same resources.

Since serverless environments rely heavily on serverless functions, in this paper we explore the threat associated with serverless function hijacking and demonstrate how 
it may inflict damage on the CIA triad of confidentiality, integrity, and availability. 

\subsection{Attacks on integrity - permission misuse}
Permission misuse, which is a deliberate and malicious attack pattern, occurs when a user's (either a human user or a system resource) permission is used inappropriately. 
In many cases, misusing permissions is possible due to excessive privileges being granted to resources and functions in violation of the least privilege principle~\cite{gill2022least,marin2022serverless}.
These over-privileged functions allow attackers to manipulate execution flows to their advantage.
For example, when permissions are granted too broadly, a function may be able to access resources or perform actions it should not be allowed to perform~\cite{pusuluri2022taxonomy}. 
In this paper, we consider permission misuse as any action that deviates from the intended utilization of a serverless function's permissions.

\subsection{Attacks on confidentiality - data leakage}
In general, data leakage is the exposure of sensitive data outside a specific network. 
Data sources are an integral component of serverless applications, and any activity involving a data source can initiate the execution of functions, promoting the exchange of data between them. 
This behavior is a result of the event-driven nature inherent to the serverless paradigm.
This, along with misconfigured resources, can lead to the leakage of confidential data.
For example, one of the most common data leakage scenarios in AWS-based environments occurs when an \emph{Amazon Simple Storage Service (S3)} bucket is misconfigured, and its permissions are set as public, thereby allowing public access to internal private data~\cite{eddy_2022,seals_seals}.
In this study, we examined the act of writing sensitive and confidential information to a public access bucket performed by a compromised serverless function. 

\subsection{Attacks on availability - denial-of-wallet}
A denial-of-wallet (DoW) attack targets cloud applications in general and serverless applications in particular. 
Similar to a typical DoS attack, a DoW attack attempts to exhaust a specific resource.
In this case, the resource is the budget allocated by the cloud administrator for computational resources. 
Accordingly, the attacker seeks to over-utilize cloud-based resources, which results in higher cloud-use bills, exceeds the allotted budget, or even pushes the application owner into bankruptcy~\cite{pusuluri2022taxonomy}.
DoW attacks take advantage of the scalability of serverless applications to artificially increase cloud resource consumption, which may eventually inflate the application's cost and exhaust its budget~\cite{kelly2021denial}. 
DoW attacks have become a significant threat to serverless applications, requiring consideration, detection, and mitigation. 
In this study, we examined cases where a function excessively consumes cloud resources.

In Section~\ref{sec:ImplementedAttacks}, we describe our implementation of permission misuse, data leakage, and DoW attacks using compromised serverless functions and demonstrate how our proposed method can detect such attacks.

\section{Background} \label{sec:background}
In this paper, we focus on the following elements of serverless computing:

\begin{itemize}[leftmargin=*]
    \item \textbf{Serverless functions} - Event-driven computing services that allow code to be run without provisioning or managing servers and automatically manage the computing resources required by that code.
    \item \textbf{Data resources} - Cloud services that are responsible for data storage.
    \item \textbf{Trigger events} - 
    A Group of services that trigger and orchestrate the activation of serverless functions by leveraging the serverless applications' event-driven nature (e.g., AWS \emph{EventBridge}, \emph{API Gateway}, and \emph{Step Function}).
    \item \textbf{Monitoring tools} - Services that audit and monitor the activities of the different entities in the cloud environment. 
    These tools can monitor both interactions between different cloud entities and the inner flow of serverless functions.
\end{itemize}

Our evaluation was performed using applications we deployed in the AWS cloud environment.
We used \emph{AWS Lambda} to implement the serverless functions, \emph{AWS Simple Storage Service (S3)} as a data storage service, and \emph{AWS DynamoDB} as a database service.
The following three trigger events were used: 
\begin{itemize}[leftmargin=*]
    \item \textbf{EventBridge} - A serverless bus event used to trigger event-driven functions (e.g., trigger Lambda functions on timed events).
    \item \textbf{API Gateway} - A service that enables the maintenance of RESTful APIs.
    In our evaluation testbed, the application receives API requests from users, and the API Gateway service triggers Lambda functions in response.
    \item \textbf{Step Function} - A workflow service used to orchestrate microservices, build pipelines using AWS services, and combine multiple Lambda functions in responsive serverless applications and microservices.
    The Step Function service allows us to orchestrate complete application workflows. 
\end{itemize}

We also use the following monitoring and application logging tools in our AWS evaluation environment:
\begin{itemize}[leftmargin=*]
    \item \textbf{CloudTrail\footnote{\url{https://aws.amazon.com/cloudtrail/}} -} A service that provides logs for every API call in the cloud environment.
    These API calls may be a database operation, an invocation of a serverless function, or any other use of an AWS service by the function.
    \item \textbf{CloudWatch\footnote{\url{https://aws.amazon.com/cloudwatch/}} - } A monitoring and observability service that collects the logs provided by CloudTrail.
    \item \textbf{X-Ray\footnote{\url{https://aws.amazon.com/xray/}} -} A debugging tool for AWS services which can be utilized in both development and production settings. 
    It collects detailed information about application flows' activities, tracks the relationships among different components, and builds an execution graph for each session. 
 \end{itemize}

\section{\label{sec:related}Related Work}

In this section, we review recent work in two areas: threat detection methods in serverless environments and anomaly detection models that use logs as their data source.

\subsection{Threat detection in serverless environments}
Several approaches for serverless security have been explored in previous research (a comparative analysis of these works is presented in Table~\ref{tab:related_work_table}).

Obetz et al.~\cite{obetz2019static} proposed a static analysis technique for extended call graphs that considers the relationships among functions and other backend services. 
The information that can be obtained through static analysis is, however, limited and covers very few of the possible threats.
In several studies, dynamic analysis techniques were suggested to better understand the behind-the-scenes activity that takes place when an application runs. 
Some focused on an application's data flow~\cite{alpernas2018secure,datta2020valve}, while others focused on an application's control flow~\cite{jegan2020guarding,sankaran2020workflow,jeganguarding}.

Alpernas et al.~\cite{alpernas2018secure} introduced Trapeze, a dynamic information flow control model designed to monitor and track the global information flow in JavaScript serverless applications.
Their solution, however, mandates adjustments to all utilized services within the application, presumes specific programming language choices for serverless functions, and requires that developers configure the information flow policies.
Datta et al.~\cite{datta2020valve} presented Valve, a platform for dynamic information flow control that uses network-level tainting to monitor functions' activity and detect attacks. 
Valve and Trapeze both rely on the collaboration of third-party services to track and monitor the functions' activity, which 
can raise concerns related to dependency and compliance, and is not feasible with all serverless architectures.

SecLambda, an architecture for securing serverless applications, was introduced by Jegan et al.~\cite{jegan2020guarding}. 
Their solution provides an extension that monitors and blocks requests within each serverless runtime environment in addition to a control program that monitors the system.
Sankaran et al.~\cite{sankaran2020workflow} proposed WILL.IAM, an access control flow model for serverless applications, which creates workflow-based policies to enforce. 
The solutions mentioned above focus on signatures of previously observed legitimate flows and specific patterns of serverless applications, without aiming to obtain any knowledge of more sophisticated patterns.
Contrary to this, our objective is to identify a wide variety of anomalies using deep neural networks' ability to learn hidden behavioral patterns in applications' activities.
By doing so, we aim to detect both known and unknown attacks with no need for prior knowledge of their signatures.
Furthermore, the proposed solutions primarily adopt preventive measures to address attacks in cloud-based environments, by reducing the attack surface through the enforcement of stringent flow and access control policies, which involve rejecting suspicious access requests.
In contrast, our approach serves as a final line of defense, granting users the ability to assess the detected anomalies' severity and determine the proper response to them to gain a better security position.

Datta et al.~\cite{datta2022alastor} presented ALASTOR, a framework that collects the system and application layers' information regarding functions’ activity. 
A global data provenance graph is then created based on the information gathered.
Kalium, an extensible security framework to enforce control-flow integrity for serverless functions and serverless applications, was proposed by Jegan et al.~\cite{jeganguarding}.

Multiple previously proposed solutions~\cite{datta2020valve,sankaran2020workflow,datta2022alastor,jegan2020guarding,jeganguarding} were implemented in a non-managed serverless environment, thus not meeting the need for threat detection solutions in a fully managed serverless environment.
In a non-managed serverless architecture, the organization is responsible for infrastructure management. 
This gives the organization greater control of the infrastructure and allows it to modify the infrastructure based on its security needs.
In this paper, we show that threat detection can be achieved in a fully managed serverless environment, using only the monitoring and logging tools provided by the cloud provider, without the need to modify the existing infrastructure.

\subsection{Log analysis for anomaly detection}
Logging and monitoring system activities are vital for evaluating a system's state and performance, and obtaining a complete picture of the activities in it. 
The ability to debug, identify root causes, and detect anomalies relies on the availability of detailed information about the system's activities.
Automated analysis of log files enables the detection of critical incidents, like system failures, at an early stage. 

Anomaly detection within log files is a broad field of study. 
Examples of  Previous research in this domain have utilized conventional machine learning techniques, such as SVM~\cite{duessel2020tracing} and clustering~\cite{liu2018integrated}, as well as deep learning methods, such as LSTM~\cite{du2017deeplog,zhao2021lstm,zhang2019robust}, autoencoders~\cite{zhou2017anomaly}, transformers~\cite{guo2021logbert}, and graph embedding~\cite{liu2019log2vec}. 
Other studies modeled logs generically~\cite{du2017deeplog,tuor2018recurrent} or modeled specific activities like network traffic~\cite{hwang2020unsupervised}, user activity~\cite{sharma2020user}, or other domains~\cite{habler2018using}.

With the growth in log volume, traditional anomaly detection methods that rely on manual inspection have become impractical. 
Manually crafted signatures and rule-based approaches that search for predefined keywords in logs exhibit limited flexibility and are unsuitable for unknown scenarios that are not pre-identified.
Self-learning anomaly detection methods can identify anomalous log events by recognizing patterns within log files without manual intervention or pre-modeling of abnormalities.
This issue has been addressed through extensive research using deep-learning neural networks~\cite{landauer2023deep}.  

Deep learning methods have also been shown to perform better than conventional anomaly detection methods when it comes to learning feature representations and using them to detect anomalies in real-world applications~\cite{landauer2023deep,pang2021deep}. 

Our model also employs anomaly detection in log data via a deep learning approach. 
However, we customized our data preprocessing specifically for the task of analyzing serverless application logs, taking into account their unique characteristics
(e.g., combining log files from various sources that contain different details with different granularity levels about a specific request, so we can gather as much information as possible, and use it to generate new features for our model).

\begin{table*}[h]
\caption{Comparative analysis of related work solutions}
\renewcommand{\arraystretch}{1.05}
\label{tab:related_work_table}
\resizebox{\textwidth}{!}{%
\begin{tabular}{|c|c|c|c|c|c|c|c|}
\hline
\textbf{Reference}  & \textbf{\begin{tabular}[c]{@{}c@{}}Non-managed \\ environment\end{tabular}} & \textbf{\begin{tabular}[c]{@{}c@{}}Dynamic \\ analysis\end{tabular}} & \textbf{\begin{tabular}[c]{@{}c@{}}Doesn't require \\3rd-Party \\ collaboration\end{tabular}} & \textbf{\begin{tabular}[c]{@{}c@{}}No infrastructure \\ modifications \end{tabular}} & \textbf{\begin{tabular}[c]{@{}c@{}}Compatibility with \\ Cloud providers' \\ native logging\end{tabular}} & \textbf{\begin{tabular}[c]{@{}c@{}}Language- \\ agnostic \end{tabular}} & \textbf{\begin{tabular}[c]{@{}c@{}}Detective \\ approach\end{tabular}} \\ \hline

Trapeze~\cite{alpernas2018secure} & \textbf{}      & \textbf{X}    & \textbf{}   & \textbf{}    & \textbf{}   & \textbf{}      & \textbf{}    \\ \hline
StaticCallGraph~\cite{obetz2019static} & \textbf{X}           & \textbf{}     & \textbf{X}    & \textbf{X}        & \textbf{X}    & \textbf{X}    & \textbf{X}      \\ \hline
Valve~\cite{datta2020valve}   & \textbf{}    & \textbf{X}     & \textbf{}   & \textbf{}    & \textbf{}    & \textbf{X}     & \textbf{}      \\ \hline
SecLambda~\cite{jegan2020guarding}    & \textbf{}      & \textbf{X}       & \textbf{X}   & \textbf{}    & \textbf{}    & \textbf{X}      & \textbf{}     \\ \hline
WILL.IAM~\cite{sankaran2020workflow}       & \textbf{}      & \textbf{X}     & \textbf{X}    & \textbf{}     & \textbf{}   & \textbf{X}    & \textbf{}     \\ \hline
ALASTOR~\cite{datta2022alastor}        & \textbf{}       & \textbf{X}     & \textbf{X}   & \textbf{}     & \textbf{}   & \textbf{X}    & \textbf{X}    \\ \hline
Kalium~\cite{jeganguarding}   & \textbf{}    & \textbf{X}     & \textbf{X}     & \textbf{}     & \textbf{}   & \textbf{X}     & \textbf{}      \\ \hline\hline
\textbf{Our Solution}       & \textbf{X}     & \textbf{X}       & \textbf{X}     & \textbf{X}        & \textbf{X}   & \textbf{X}     & \textbf{X}                                                                  \\ \hline
\end{tabular}%
}
\end{table*}

\section{\label{sec:method}Proposed Method}

In this section, we present the proposed methodology of our anomaly detection model. 
We start by introducing the notations used, then we describe our method for the compromised function detection model, and finally, we present the update component integrated into our model.

\subsection{Notations} \label{subsection:Notations}

The logs used by our threat detection model contain \emph{application flows}.
An \emph{application flow} is usually triggered by an action outside of the cloud environment (e.g., a REST API call or a file upload to a public storage service).
Each \emph{application flow} is constructed by sequences of \emph{function flows}, where each \emph{function flow} is composed of a sequence of \emph{function events} such as internal functions' calls or resource operations.

\begin{figure}[h]
    \centering    
    \includegraphics[width=\columnwidth]
    {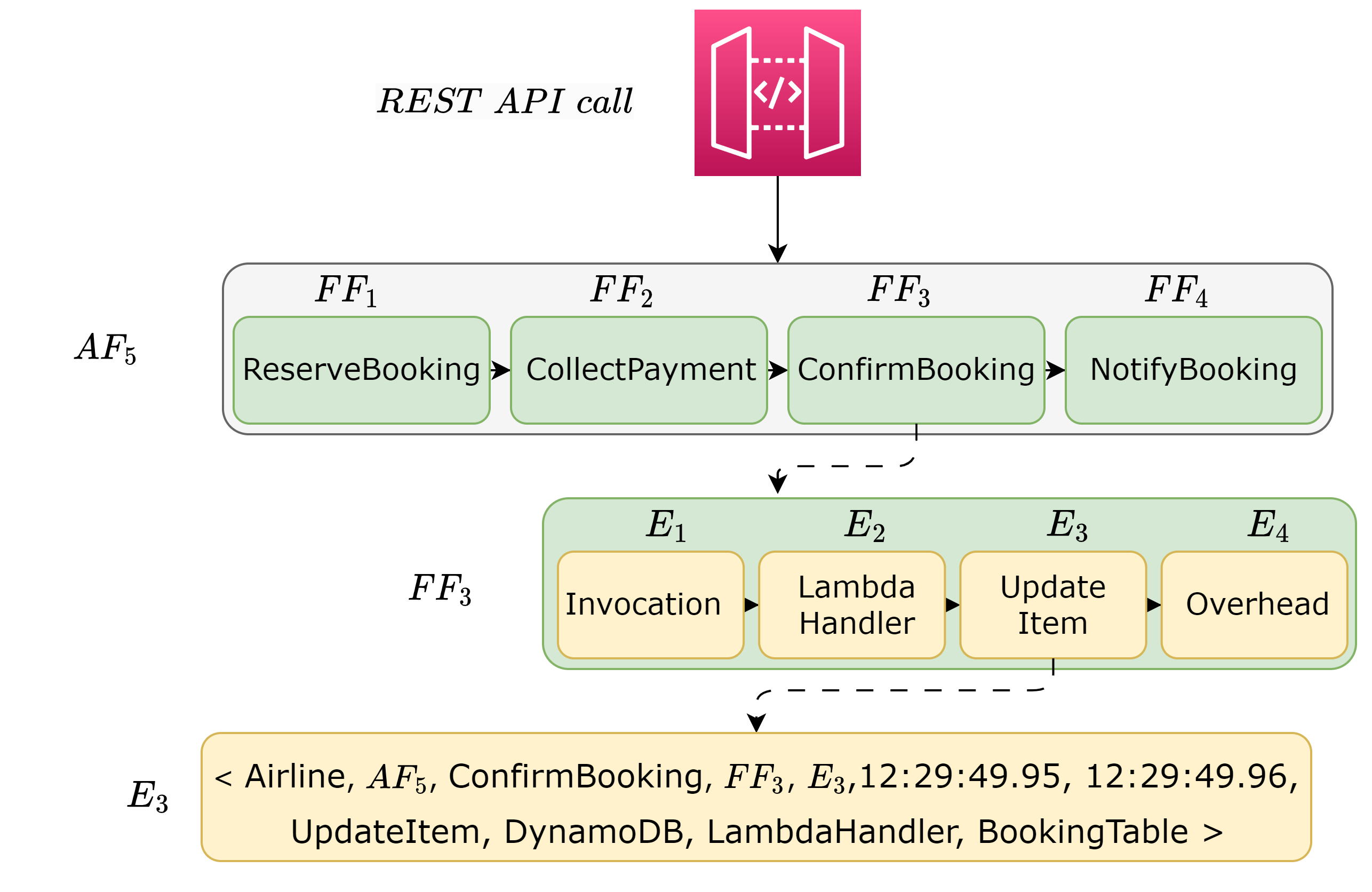}
    \caption{An \emph{application flow} of a booking process in our testbed.
        This \emph{application flow} is comprised of a series of \emph{function flows}, each containing a sequence of \emph{function events}. 
        Each \emph{function event} has a log entry that provides comprehensive details about it.}
    \label{fig:exampleForNotations}
\end{figure}

An example of an \emph{application flow} in our testbed is presented in Figure~\ref{fig:exampleForNotations} (denoted by $AF_5$). 
As can be seen in our booking application flow example, the \emph{application flow} is triggered by a REST API call and is composed of four \emph{function flows}. 
In this example, the third \emph{function flow} is the flow of a function named ConfirmBooking (denoted by $FF_3$), which contains a sequence of four \emph{function events}.

\noindent Each \emph{function event} is defined by the following properties:
\begin{itemize}
    \item \textbf{\textit{ApplicationName}} is the name of the \emph{application} in which the event occurred.
    \item \textbf{\textit{ApplicationFlowId}} is the identifier of the \emph{application flow} in which the event occurred.
    \item \textbf{\textit{FunctionName}} is the name of the \emph{Function} in which the event occurred.
    \item \textbf{\textit{FunctionFlowId}} is the identifier of the \emph{function flow} in which the event occurred.
    \item \textbf{\textit{EventId}} is the identifier of the event.
    \item \textbf{\textit{StartTime}} is the event's start time.
    \item \textbf{\textit{EndTime}} is the event's end time.  
    \item \textbf{\textit{EventName}} is the name of the operation that took place 
    (e.g., initialization, invocation, overhead, a resource operation (GetItem, PutItem, etc.), or the name of an internal function that was called).
    \item \textbf{\textit{EventType}} is the type of event (e.g., an internal function call, the event's serverless service name (DynamoDB, S3, SNS, etc.), or other).
    \item \textbf{\textit{EventParentName}} is the name of the event's predecessor.
    In some cases, an event may be triggered by another event. 
    In such cases, the log provides information about the calling event. 
    For example, in Figure~\ref{fig:exampleForNotations} the ``Update Item" event ($E_3$) is called by the ``Lambda Handler" event ($E_2$).
    \item \textbf{\textit{EventTargetResource}} is the event operation's target resource.

\end{itemize}

\subsection{Compromised function detection model} \label{subsection:Compromised function detection model}

\begin{figure}[h]
    \centering
    \includegraphics[width=0.99\textwidth]{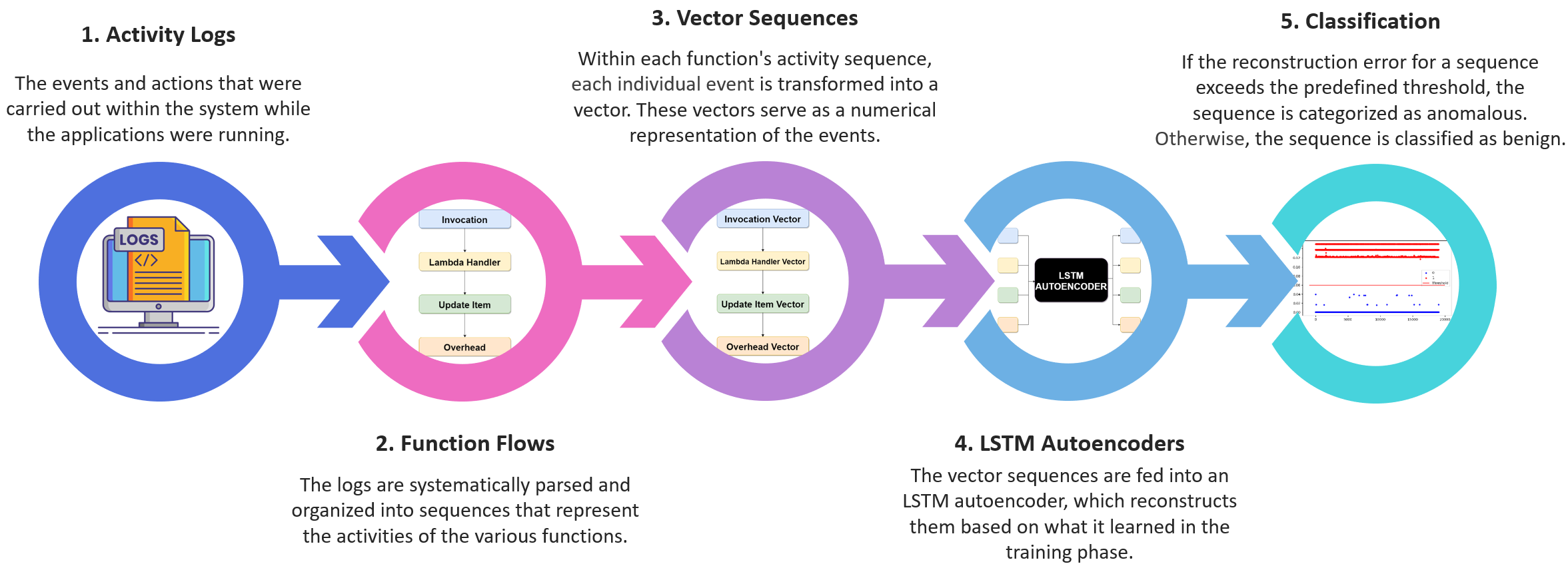}  
    \caption{An overview of the essential processes in our compromised functions anomaly detection model.}
    \label{fig:models_steps}
\end{figure}

The compromised function detection model focuses on serverless functions' flow and aims to detect deviations from their normal activity. 
The events that occur during the execution of each function are represented by a sequence of vectors, where each vector represents an event in the sequence.
This way we are able to use the high granularity of the cloud provider's logs to model and detect abnormalities in the order of the events in a function's execution as well as in the events' attributes.
This is done by training several LSTM autoencoders.
The LSTM autoencoder is a deep learning architecture that learns to encode and decode sequences. 
Each autoencoder processes the sequences of the event vectors, capturing temporal dependencies and complex patterns present both within the data and in each event within these sequences. 
Ultimately, it identifies anomalous sequences based on the insights it has gained. 
Figure~\ref{fig:models_steps} provides an overview of the key steps in the model's workflow.

\subsubsection{\textbf{Data collection}}
We use the cloud provider's native hyper-logging and monitoring tools to collect our data.
The platforms of all major cloud providers include such tools.\footnote{\url{https://cloud.google.com/free/docs/aws-azure-gcp-service-comparison}} (i.e., AWS, Microsoft Azure, Google Cloud) 
We collect the raw logs containing information regarding the activity of the serverless applications implemented and use them to train our compromised function detection model under the assumptions that (1) the attacker is unable to control logging, and (2) the logs are unmodified and reflect the real activity of the system (i.e., functions).

\subsubsection{\textbf{Preprocessing}}
As described in Section~\ref{subsection:Notations}, the data we collect consists of $application flows$ representing the activities of the examined applications in a serverless environment.
From these $application flows$, we use the $function flows$ of the examined functions. 
In order to represent each function execution, in the preprocessing phase, we create a sequence of $n$ vectors. 
Each vector represents an event that occurred as part of the $function flow$.
The relevant features of each event are extracted from the log entry associated with the event, and new features are generated based on these features. 

\noindent The features extracted directly from the event's log entry include (see Section~\ref{subsection:Notations}): 
\emph{ApplicationName, EventName, EventType, EventParentName}, and \emph{EventTargetResource}. 

\noindent The new features generated based on the event's log entry include:
\emph{EventDuration} - the difference between the event's \emph{StartTime} and \emph{EndTime}; \emph{RelativeStartTime} - the time that passed since the \emph{StartTime} of the first event in the sequence; and the depth of the event - the depth of the parent event plus one or one if the event has no \emph{EventParentName}.

\noindent These features are used to construct the event vectors.

\begin{figure}[h!]
    \centering
   \includegraphics[width=0.95\columnwidth]{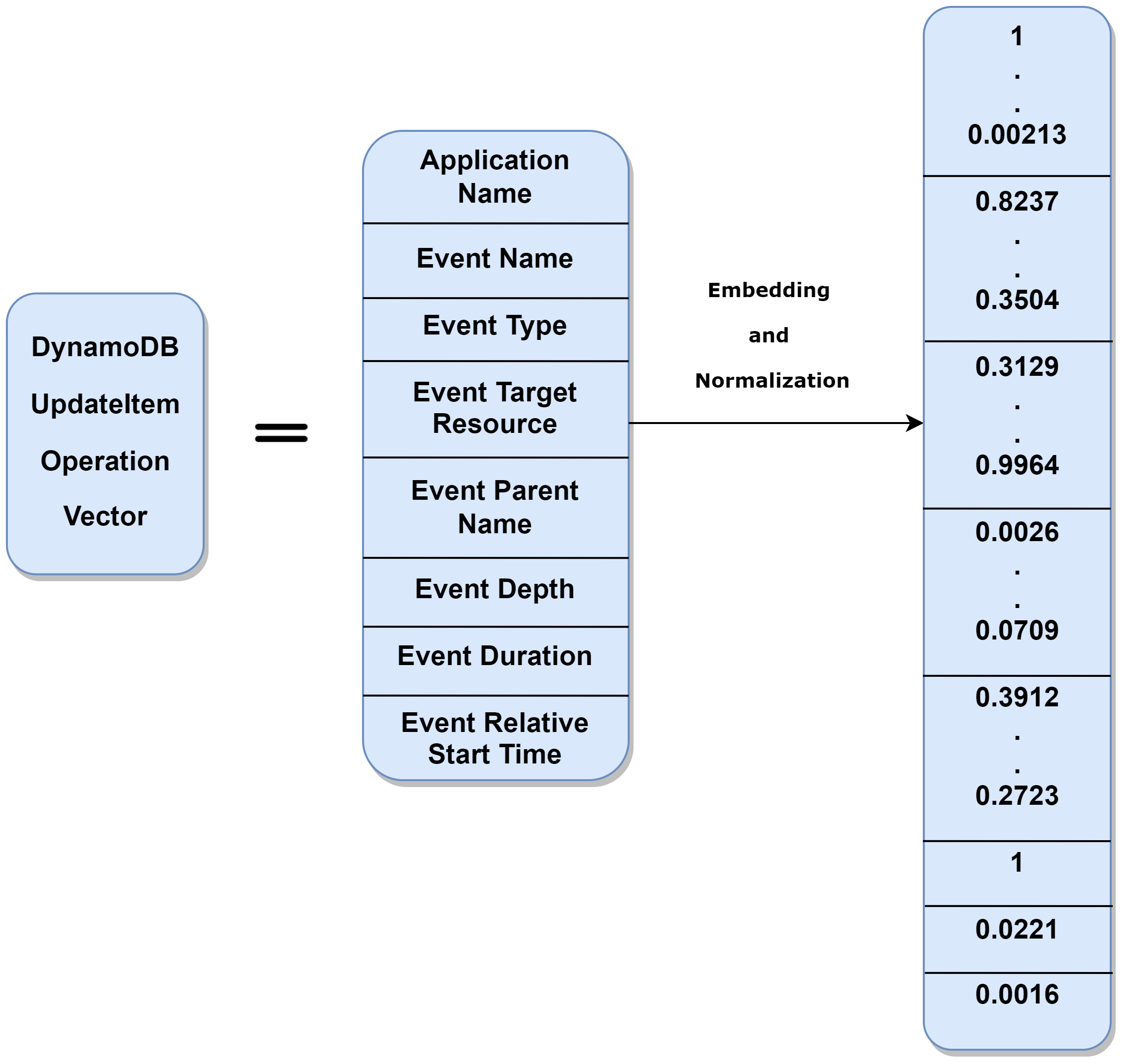}  
    \caption{Example of an event vector associated with the $UpdateItem$ operation.
    Each categorical feature is represented by a pair of numerical values, and subsequently, all the values in the vector are normalized to fit within the [0,1] range based on their respective minimal and maximal values.}
    \label{fig:update_item_event_vector}
\end{figure}

In order to represent categorical features, we use the Chars2vec library,\footnote{\url{https://github.com/IntuitionEngineeringTeam/chars2vec}} which embeds words using a pretrained model (eng\_50) and returns a vector of 50 dimensions for each word.
To represent each word, we use the first four dimensions of the vector. 
Our vocabulary is primarily comprised of the names of serverless services, resources, and operations. 
Consequently, despite the model's potential use across numerous applications and functions, the vocabulary will consistently encompass a finite number of unique words. 
As a result, the use of the first four dimensions for each vector effectively gives distinct meanings for each word within the limited vocabulary.
In the training phase, the minimal and maximal values of each feature are stored. 
Using these values, we employ min-max normalization to the feature values within both the training and test sets.
Figure~\ref{fig:update_item_event_vector} presents an example of an event vector associated with a $UpdateItem$ operation in DynamoDB, before and after it was encoded and normalized. 

\subsubsection{Anomaly detection}
In order to learn the normal behavior patterns of the serverless functions, we train an LSTM autoencoder using the vector sequences we created as training samples. 
Using the sequences generated in the preprocessing stage as input to the LSTM autoencoders, each autoencoder learns the typical patterns of the functions' sequences.
The LSTM autoencoder encodes and decodes each function instance represented by a sequence of vectors.
Then, we use the reconstruction error (computed using the mean square error (MSE)) of each sequence to determine the anomaly score of the instance it represents.
Sequences with a reconstruction error exceeding the predefined threshold are marked as potential attacks, suggesting that the observed activities deviate from the expected norm. 
Conversely, if the reconstruction error falls below the threshold, the sequence is classified as benign, indicating that the activities align with the anticipated~behavior.

We trained three distinct autoencoders, each customized for a specific sequence size.
We handle vector sequences whose length differs from the autoencoder's input layer length (referred to as the window size) as follows:
\begin{itemize}
    \item If the event sequence of a function exceeds the window size, we generate new sequences by applying a sliding window technique with a stride of one. 
    For instance, if the window size is three and the function sequence contains four events, we create two new sequences. 
    The first sequence consists of the initial three events from the function sequence, and the second sequence is comprised of the last three events from the function sequence.
    \item If the event sequence of a function is shorter than the window size, the sequence is padded with zero vectors.
\end{itemize}
To identify the optimal window size for detecting attacks, we performed a comprehensive grid search. 
This analysis revealed that different attacks were effectively captured within distinct timeframes. 
The most favorable results were achieved using window sizes of 3, 5, and 10.
The architecture of the network is identical for all of the autoencoders and is described in Section~\ref{sec:evaluation}.
An event sequence is classified as anomalous if at least one of the autoencoders classified it as such.

The classification threshold for categorizing each sample as either benign or anomalous is determined based on the reconstruction errors observed in the validation set during the training phase. 
Our objective is to establish a threshold that strikes a balance, ensuring it is neither overly strict nor overly lenient.
To achieve this balance, we apply the DBSCAN algorithm to the validation set errors. 

DBSCAN is a density-based clustering algorithm designed to identify dense clusters within the data, differentiating them from regions of lower density~\cite{dbscan}.
This approach enables us to base the threshold value solely on the normal samples while disregarding the outliers in the dataset.
The DBSCAN algorithm relies on two essential parameters: epsilon and minimum samples. 
Epsilon determines the maximum distance between two points for them to be considered part of the same cluster. 
The minimum samples parameter dictates the minimum number of data points required to form a cluster.
The algorithm's epsilon parameter is set to the 99th percentile of the validation set's reconstruction errors, and the minimum sample size is set at five.
In the process of eliminating outliers, we focus on clusters that comprise more than 5\% of the total validation set size. This criterion helps remove noise and spurious clusters, ensuring that only meaningful clusters are retained for analysis.
Finally, to determine our threshold, we set it as double the highest reconstruction error within the remaining clusters.

\begin{figure}[h]
    \centering
    \includegraphics[trim={0 0 0.5cm 0.9cm},clip,width=0.99\columnwidth]{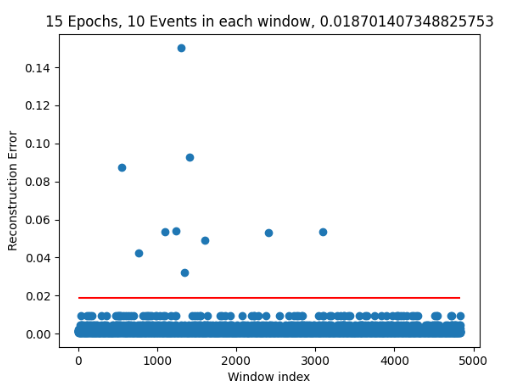}  
    \caption{An illustration of the errors observed in the validation set of an autoencoder with a window size of 10; the red line indicates the threshold (0.0187) established based on these errors.}
    \label{fig:validation_set_errors}
\end{figure}

Figure~\ref{fig:validation_set_errors} illustrates the errors observed in the validation set of one of the autoencoders, and the threshold established based on them through the described procedure.

\subsection{Updating the model} \label{subsection:Model Update}

Understanding the significance of a threat detection tool's ability to adapt to evolving circumstances and its need to stay current with the latest information, we have integrated a model update component into our framework. 
This component serves the following purposes:
\begin{itemize}
    \item Incorporating new functions. As organizations maintain and develop their applications, they frequently introduce new functionalities. Our model needs to stay aligned with these developments.
    \item Managing code modifications and concept drifts.
    As the application evolves, code updates enhance features and resolve issues, while concept drifts occur due to changing data patterns influenced by user behavior, market dynamics, or external factors.
    To ensure the model's effectiveness, it must seamlessly adapt to these changes.
    \item Enhancing the model's capability to identify rare legitimate events. This ensures that genuine occurrences are not mistakenly flagged as threats.
\end{itemize}

Updating the entire model can be a resource-intensive task, requiring significant computational power and time. 
In recognition of this challenge, our model update component offers a more efficient approach through iterative updates (delta updates). 
This way, only the necessary adjustments or changes to the model are implemented, rather than retraining the entire model from scratch.

In this procedure, we fine-tune the model using the new data along with approximately 10\% of the original training set used to train the original model. 
We found this combination of training data and new data yields the best results. 
It is worth noting that the data from the original training set which is employed when updating the model, may be deliberately chosen.
This planned selection approach can be tailored to account for outliers, such as rare legitimate events identified using the original model, or to accommodate outdated data, such as refactored and modified functions.
We discuss this point further in the evaluation section.

To establish the new threshold, the entire original validation set is retained, and the new validation data is added to it.
To determine the new threshold, we predict the errors of the combined validation set using the updated model and subsequently apply the original automated threshold procedure to calculate the adjusted threshold.

\section{Serverless Cybersecurity Evaluation Testbed} \label{sec:testbed}

To analyze and assess our model's performance, we required an experimental serverless cloud environment.
For this purpose, we created a serverless cybersecurity evaluation testbed, designed to achieve the following goals: 
(1) cover various serverless development paradigms employing a wide range of serverless cloud resources, 
(2) generate a large amount of authentic log data reflecting real-life activities, and 
(3) simulate the consequences of typical attack scenarios.
We created our testbed on the AWS platform, and employed the AWS X-Ray service as a native fine-grained logging tool.

\subsection{Testbed applications}
\begin{figure}
    \centering
    \subfloat[\centering Airline Booking Application]{{\includegraphics[width=0.95\textwidth]{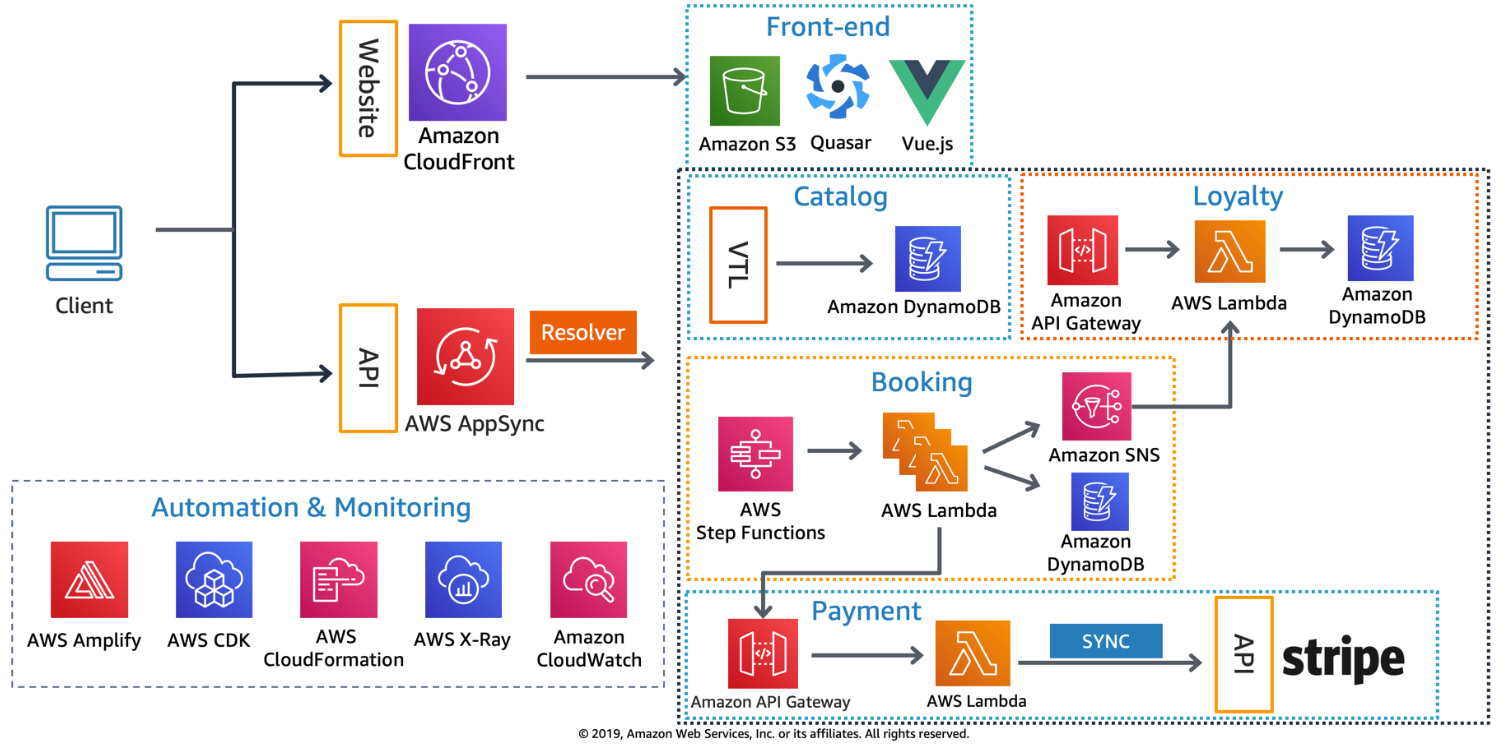} }}%
    \qquad
    \subfloat[\centering VOD Application]{{\includegraphics[width=0.95\textwidth]{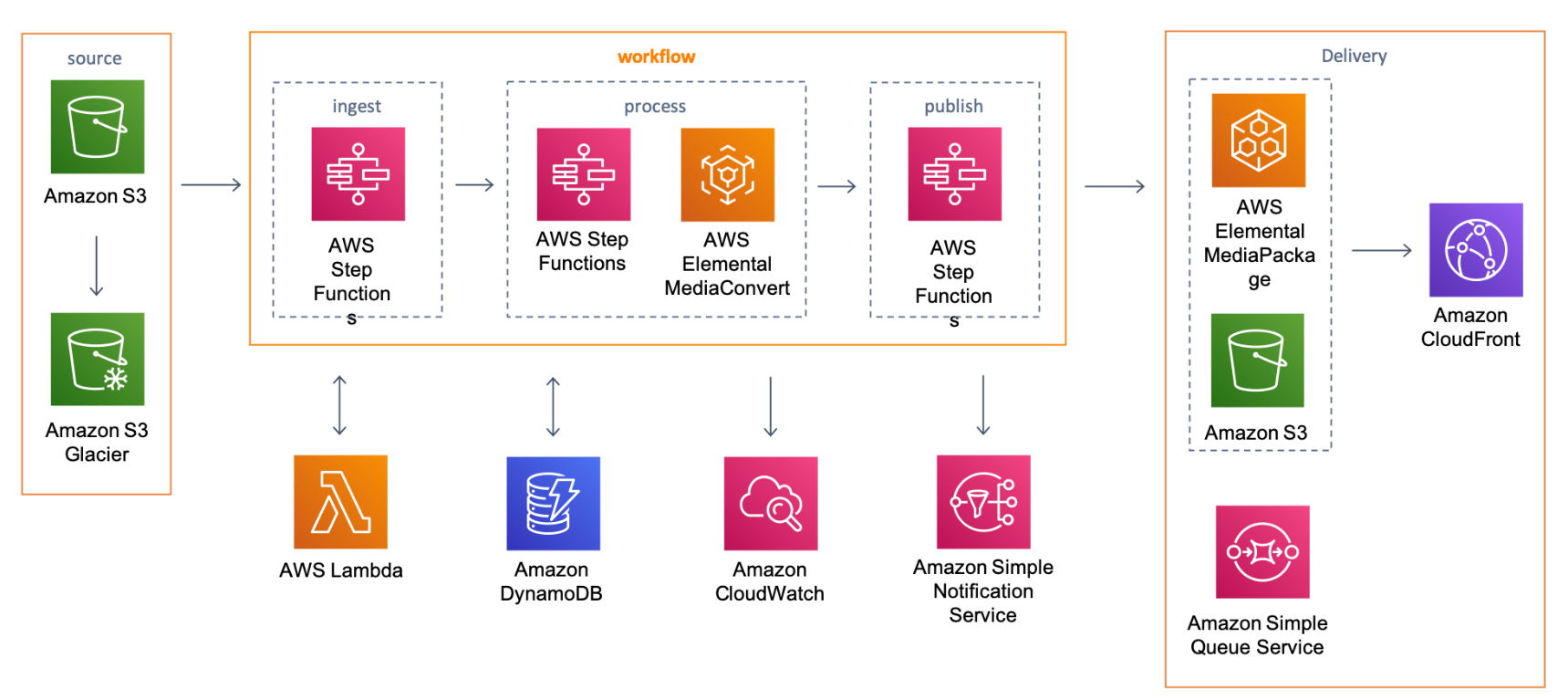} }}%
    \caption{An overview of the architectures of the testbed applications, highlighting the interactions and processes facilitated by various AWS services employed by each of these applications \cite{airline_github, vod_github}.}%
    \label{fig:apps_architecture}%
\end{figure}

The testbed consists of two different applications:
(1) an airline booking application,\footnote{https://github.com/aws-samples/aws-serverless-airline-booking} and 
(2) a video on demand (VOD) application.\footnote{https://github.com/aws-solutions/video-on-demand-on-aws}
These applications were selected, because they represent two different serverless development approaches. 
They also provide an opportunity to explore a wide range of AWS serverless services in the testbed.
In the airline booking application, each serverless function serves a specific, well-defined purpose in the workflow. 
Conversely, in the VOD application, the functions have a more versatile and generic nature and are capable of being used in various contexts and scenarios.
The architectures of the two applications are presented in Figure ~\ref{fig:apps_architecture}.

\subsubsection{Airline booking application}
This application is comprised of multiple REST API endpoints that facilitate user interaction. 
It leverages various AWS services, including four DynamoDB tables, four S3 buckets, SQS and SNS services for queuing and notifications, the AppSync API service, and 12 Lambda functions. 
Notably, some of these Lambda functions are coordinated using the AWS Step Function service.

This application encompasses three main types of business processes:
\begin{itemize}
\item Data Retrieval: This type of business process involves API requests that retrieve information from a DynamoDB table and subsequently deliver the data to the user. 
These events are characterized by their brief duration and involve single accesses to data sources within the system.
\item Booking Process: In this process, an API request to book a flight initiates a sequence of data access operations and Lambda function invocations that exchange information. 
These events are orchestrated by a step function, ensuring the coordination of various steps.
\item Time-Based Events: The third type of business process is triggered by AWS EventBridge, which initiates specific actions at predetermined time-based intervals.
\end{itemize}

\subsubsection{VOD application}
As a core part of its functionality, this application plays a role in decoding videos into a streaming format and storing them in a public S3 bucket. 
It utilizes three S3 buckets, two DynamoDB tables, and 17 Lambda functions. 

In contrast to the airline booking application where only one of the business processes is orchestrated by a Step Function, the VOD application employs four Step Functions, each of which is responsible for a distinct core business process.
The three processes in the main video format conversion workflow are as follows:
\begin{itemize}
\item Ingest: verifies the validity of the uploaded video files.
\item Process: converts the video files from their original coding format to a different format using the AWS Elemental MediaConvert service.
\item Publish: generates a link to the file's location in the cloud and dispatches it via email to a predefined email address using the AWS Simple Notification Service (SNS).
\end{itemize}
The fourth process is added to the original application's code by us and does not appear in Figure~\ref{fig:apps_architecture}. 
It is dedicated to handling video uploads through emails and utilizes the AWS Simple Email Service (SES).

Users can interact with the application in two ways: manually uploading files to the source S3 bucket or sending emails to a predefined mail endpoint. 
Once a file is added to the source S3 bucket, the subsequent processes are automated via EventBridge triggers and Step Functions.

Each application is equipped with its own dedicated DynamoDB table, which contains 
configuration information for the simulation process (mainly the probabilities associated with each execution in order to run anomalous code).

\subsection{Data simulation}

In our testbed environment, we strive to generate experimental data that faithfully replicates the behavior of real-life applications. 
To simulate genuine user actions, we developed Python scripts that execute functions in various sequences, mimicking users who follow different paths and perform distinct operations.
Furthermore, we simulate potential errors that could arise during the executions carried out in these varied pathways.

\subsubsection{Airline booking application}
Our simulation involves traffic through the airline booking application's frontend website and incorporates fixed probabilities associated with different types of booking reservations (i.e., one-way or two-way flights).
To elaborate:
\begin{itemize}
\item There is an 80\% probability of booking two-way tickets, which involves reserving two tickets with opposite source and target airports. 
The remaining 20\% is allocated for checking the booking page and ordering one-way tickets.
\item During the booking process, there is a 70\% probability of navigating to the profile page to obtain a response from the ``get loyalty" lambda function.
\item The simulated user can visit two different pages: the loyalty page and the booking page. 
Each visit triggers a distinct REST API call which, in turn, activates the APPSync API service in the cloud environment.
\item When booking tickets, there are two options: one-way or two-way bookings. 
For one-way bookings, a random airport is selected from a predefined list as the starting location, and another random airport from the list is chosen as the target location.
\item In the case of two-way bookings, the first ticket's starting location is set at a specific airport, while a random airport from the list is selected as the target. 
For the second ticket, the roles are reversed, with the same airports used in the opposite roles.
\end{itemize}

This simulation process generates various booking scenarios, including different types of reservations and the associated API interactions.

Completion of the process of booking a flight in the simulator triggers the step function within the cloud environment, initiating the step function's workflow. 
To ensure that a wide range of activity patterns are seen in the testbed, this workflow incorporates different error-handling mechanisms that activate various combinations of Lambda functions.
The likelihood of these paths being executed is determined by the configuration table of the airline booking application.

To mimic genuine user activity within the application, we insert time intervals that introduce time gaps between users' actions, taking into account the time of day and the specific actions users tend to perform. 
This approach produces a distributed activity load that closely resembles the patterns seen in actual user behavior.

\subsubsection{VOD application}
In the simulation of the VOD application, we established a staging S3 bucket that functions as a virtual local drive" for users, enabling them to upload video files to the application's source bucket.
To perform cloud operations that trigger the application's workflow, we employ the AWS SDK for Python, commonly known as Boto3. 
This facilitates actions such as file uploads to the S3 buckets.
For the mailing process that was added by us, we configured Amazon SES to receive and store emails originating from our Amazon Route 53 service domain. 
This setup ensures that the mailing flow is properly initiated.
In addition to the direct uploads facilitated by Boto3, the simulator also employs Python's email library to send emails to our domain, adding another layer of functionality to the simulation process.

To simulate realistic timing for file uploads in the VOD application simulation, we leverage the Trending YouTube Video Statistics dataset from Kaggle\footnote{\url{https://www.kaggle.com/datasets/datasnaek/youtube-new}}. 
This dataset provides information about the upload times of popular videos in the United States, serving as the basis for simulating activity in the VOD application.
We employ the distribution of video uploads from this dataset in order to model the activity patterns in the VOD application:
\begin{itemize}
\item{There is an 80\% probability that a standard "put object" operation will be initiated by an S3 upload, while there is a 20\% chance of a mail upload, unless the mailing service has already handled 200 emails, in which case it reaches the defined limit.}
\item{In the case of a regular upload, there is a 90\% chance of a successful upload, while there is a 10\% chance that the format name will be altered to a format that the MediaConvert service cannot process.}
\item{Mail uploads are guaranteed to be successful.}
\end{itemize}

This approach ensures that the simulation closely mimics real-world upload scenarios, incorporating factors such as the upload type and success rates.

\subsection{Attacks implemented } \label{sec:ImplementedAttacks}

To demonstrate the feasibility of our proposed method, we implemented several attack scenarios in the testbed which demonstrate common threats in the serverless threat landscape.
These attacks illustrate the effects that real attacks can have on a serverless application, including the artifacts they leave behind in serverless activity logs.

\subsubsection{Attack on integrity - permission misuse}
In this scenario, we simulate an attack that modifies the serverless function's execution flow. 
Specifically, we simulate the abuse of permissions by a serverless function that was granted permission to access and operate on a data source.
In this case, the permissions are misused in order to change the context in which the function operates for the benefit of the attacker.
In the first version of this scenario, the order of the data operations in the function's event sequence is different than usual (e.g., instead of updating an item and then obtaining an item from a table, the function performs a get item operation and then updates that item). 
In the second version of the attack, the function performs a different operation than the one it usually performs (e.g., updating an entry in a table instead of writing a new entry to it).
The third version involves adding an additional action to the function's workflow (e.g., additional access to a DynamoDB table).

\subsubsection{Attack on confidentiality - data leakage}
In this particular attack scenario, the malicious actor leverages the serverless function to illicitly exfiltrate sensitive information to a publicly accessible data source.
To achieve this, the attacker enhances the function's customary workflow by incorporating functionality that directs it to transfer private data to an external AWS S3 bucket.

\subsubsection{Attack on availability - denial-of-wallet}
Two versions of this attack scenario are simulated.
In the first version, the function repetitively invokes the same API call as part of its execution, resulting in multiple accesses to a data source. 
Consequently, both the data transported and the function's runtime experience an increase.
The second version of the attack is designed to leave fewer tracks compared to the first. 
In this version, the runtime of some of the functions is increased by just a few seconds (4-5).
This might not raise suspicion if it is a one-time occurrence, but if numerous functions execute for an extended duration, they will consume more resources and ultimately increase the cost to the user.\\

These attacks are implemented in the main workflow of the applications examined in this study.
For each application, we chose specific functions and ``infected" them with the attack scenarios.
During data collection, we configured 10\% of the instances of each ``infected" function to run the malicious code and logged the anomalous instances in order to label each of the instances in the test set as benign or anomalous.

\section{\label{sec:evaluation}Evaluation} 

In this section, we evaluate our model's performance. 
We introduce our dataset statistics, describe the network architecture and hyperparameters used in our solution, present the results for the primary model and the additional update component, and discuss their implications.

\subsection{\label{sec:Dataset}Dataset} 

To evaluate the performance of our compromised function detection model, we conducted simulations of the activities of two distinct applications, as outlined in Section~\ref{sec:testbed}.

The execution time of the VOD application simulator is directly derived from the number of files uploaded and processed. 
In each of the phases, the training and testing, we uploaded 3,000 files, which resulted in an approximate six-hour runtime for each set of simulations.
The execution time of the airline booking application simulator is derived from the number of iterations the simulator is required to perform. 
In each of the phases, the training and testing, we configured the simulator to run for 800 iterations, resulting in an estimated runtime of around four hours for each set of simulations.

For each function, we established a consistent anomaly rate of 0.1. 
Then, each function invocation in the test set was chosen at random to execute either the anomalous or benign code, in accordance with this predetermined rate.

Our evaluation focused on the functions with the most prominent roles in each application, a total of 12 functions — eight from the Airline Booking application and four from the VOD application were considered to have the most prominent roles. 
In this assessment, we simulated attacks on three functions in the Airline Booking application and all four functions in the VOD application. 

We divided the training data into training and validation sets using an 80/20 ratio.
Table~\ref{tab:data_table} provides details regarding the number of instances (vector sequences) and anomalies for each dataset of each application.
The amount of attacks simulated in each application is presented in Table~\ref{table:attacks_table}.

\begin{table}[h]
\centering
\caption{The number of attacks simulated in each application}
\renewcommand{\arraystretch}{1.0}
\label{table:attacks_table}
\begin{tabular}{|l|l|l|}
\hline
\textbf{Attack}                                    & \multicolumn{1}{l|}{\textbf{Airline}} & \multicolumn{1}{l|}{\textbf{VOD}} \\ \hline
{Data Leakage}                               & 52              & 117           \\
{Denial of Wallet -  Repeated Operation}          & 141              & 800          \\
{Denial of Wallet - Increased Duration}      & 80              & 277          \\
Permission Misuse - Change Operation Order & 72              & 265          \\
{Permission Misuse - Different Action}       & 139              & 0            \\
{Permission Misuse - Additional Action}      & 0              & 118           \\ \hline
\end{tabular}
\end{table}
\begin{table}[h]
\caption{The number of instances (vector sequences) and anomalies for each dataset of each application}
\renewcommand{\arraystretch}{1.1}
\label{tab:data_table}
\resizebox{\columnwidth}{!}{%
\begin{tabular}{|c|c|c|c|c|}
\hline
\textbf{Application}                                                & \textbf{Train Set} & \textbf{Validation Set} & \textbf{Test Set} & \textbf{Anomalies} \\ \hline
\textbf{\begin{tabular}[c]{@{}c@{}}Airline \end{tabular}} & 5,927              & 1,481                   & 4,497             & 484                \\ \hline
\textbf{VOD}                                                        & 14,568             & 3,642                   & 16,374            & 1,577              \\ \hline
\end{tabular}%
}
\end{table}

\subsection{\label{sec:hyperparameters}Network architecture, hyperparameters, and experiment details} 

\begin{figure}[h]
    \centering
    \includegraphics[width=\columnwidth]{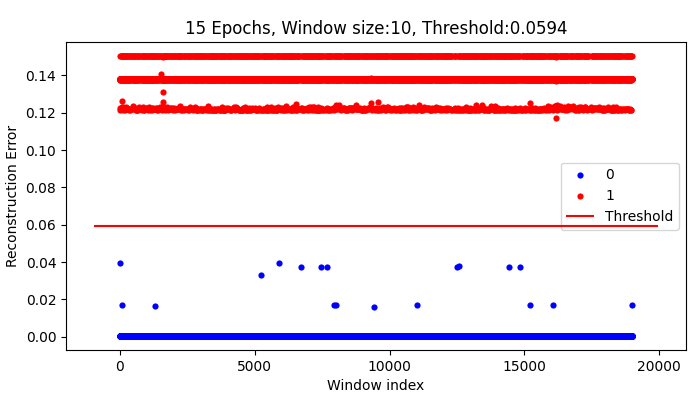}  
    \caption{A visualization of the autoencoder's output: reconstruction errors of a test set comprised of around 20,000 instances.  
   The blue dots correspond to sequences within benign function executions, while the red dots denote sequences in anomalous function executions. 
    The chosen threshold, derived from the validation set, is depicted by the red line.}
    \label{fig:TestSetErrorsExample}
\end{figure}

We conducted a grid search to select the optimal hyperparameters for the autoencoders and the model itself. 
The selection of parameters was guided by identifying the clearest delineation between the reconstruction errors of anomalous and benign instances, as derived from the outputs of each autoencoder. 
Figure~\ref{fig:TestSetErrorsExample} illustrates an optimal scenario in which benign instances exhibit minimal reconstruction errors, while anomalous instances display significantly elevated reconstruction errors. 
This clear distinction between the two types of instances results in a perfect separation.

First, we experimented with the number of hidden layers and neurons in each layer, the activation functions within each layer, and the optimization algorithm for the network.
The final network architecture we selected is identical for all of the autoencoders, and it consists of seven fully connected feed-forward hidden layers.
The encoder and decoder layers consist of 128, 64, and 32 neurons (from the outer layer to the inner one), and the activation function for these layers is ReLU. 
A middle layer containing 64 neurons replicates the inner layer of the encoder by a factor equal to the number of events in the sequence.
We used the Adam optimizer for the network, and the loss function is the mean squared error. 

We also examined factors such as the number of epochs, the batch size, the learning rate, the dimensions of the input sequences, and the threshold used for the classification of instances.
The optimal approach was training three separate autoencoders, each with a distinct sequence size: 3, 5, and 10. 
These autoencoders were trained over the course of 15 epochs, with a batch size of 32 instances, and a learning rate of 0.0001.
An instance (sequence) is classified as anomalous if at least one of the autoencoders classifies it as such.
The method for choosing the threshold for each autoencoder is described in Section~\ref{subsection:Compromised function detection model}.

When updating the model, all hyperparameters but the learning rate remain unchanged (compared to the hyperparameters used to train the original model).
This includes the original minimum and maximum values for each feature used for feature normalization.
The learning rate used in this phase is set at 0.0001.

Our experiments were performed using a computer equipped with a 4-core Intel Core i7-10510U CPU @ 1.80GHz with 16 GB of RAM.

\subsection{\label{sec:Results}Results}

The threat detection results obtained for each application and attack are presented in Table~\ref{tab:results_table}. 
Further details regarding the implementation of each attack can be found in Section~\ref{sec:ImplementedAttacks}.
The overall results obtained for both applications are presented in Table~\ref{tab:multiapp_results_table}.
Table~\ref{tab:multiapp_results_table} provides a comprehensive summary of the results for all significant functions within both applications, including those unaffected by simulated attacks, while 
Table~\ref{tab:results_table} presents the results for each attacked function.

\begin{table}[h]
\caption{Detection results for both applications}
\renewcommand{\arraystretch}{1.10}
\label{tab:multiapp_results_table}
\resizebox{\columnwidth}{!}{%
\begin{tabular}{|c|c|c|c|c|c|}
\hline
\textbf{Application}                                                            & \textbf{Presicion} & \textbf{Recall} & \textbf{F1 Score} & \textbf{FPR} & \textbf{FNR} \\ \hline
\textbf{\begin{tabular}[c]{@{}c@{}}Airline + VOD\end{tabular}} & 0.967              & 1               & 0.983             & 0.003        & 0            \\ \hline
\end{tabular}%
}
\end{table}

As evident from the tables, high values were obtained for the precision, recall, and F1 score, while the false positive rate (FPR) and false negative rate (FNR) were reasonably low for all types of attacks.

In general, the precision scores achieved for the VOD application were inferior to those of the airline booking application. 
This discrepancy can be attributed to a greater occurrence of false alarms associated with a particular function within the VOD application.
Both the \emph{permission misuse (additional operation)} and \emph{data leakage} attacks were directed at this particular function.
While there were very few cases where benign sequences were classified as anomalous, we sought to understand what was causing these misclassifications.
Our analysis showed the vast majority of them involved cold-starting function instances.
As such an event was relatively rare in the training set, the model was not sufficiently experienced with sequences that included it.
We conclude that most of such cases stem from events in the functions' executions that were not seen in the training set. 
Accordingly, we conclude that the more diverse the training set is in terms of the event sequences it contains, the more complete the model will become, and the better the results. 
Nevertheless, the model has the ability to highlight valuable log data for a manual inspection of the security analysts in both applications. 
Given the perfect recall and the low false positive rate,
we believe that our model will serve as an invaluable tool for data analysts - instead of going through all of the available logs, an analyst can focus on those log entries that are the most likely to be relevant.

It should be emphasized that the proposed method is used to model serverless functions. 
In a typical serverless application, each function generally performs a small atomic role. 
Therefore, the expected number of execution flow variations for each is limited, as is the expected number of rare events. 

Still, a model's ability to learn rare benign behaviors and adapt to changes is critical in reducing the number of false alerts.
To achieve that, our model can be retrained to accommodate legitimate changes in the code and rare benign events.
Our future work will involve utilizing imbalanced data correction methods, such as the oversampling of rare benign sequences, in order to provide the model with an improved training set and enhance its learning process.

\begin{table}[h]
\caption{Detection results for each simulated attack \& application}
\renewcommand{\arraystretch}{1.25}
\label{tab:results_table}
\resizebox{\columnwidth}{!}{%
\begin{tabular}{|c|c|c|c|c|c|c|}
\hline
\textbf{Application}  & \textbf{Attack}   & \textbf{Precision} & \textbf{Recall} & \textbf{F1 Score} & \textbf{FPR} & \textbf{FNR} \\ \hline
\multirow{5}{*}{\textbf{\begin{tabular}[c]{@{}c@{}}Airline \end{tabular}}} & \begin{tabular}[c]{@{}c@{}}Permission Misuse: \\ Different Order of Operations\end{tabular} & 0.973              & 1      & 0.986             & 0.003        & 0            \\ \cline{2-7} 
  & \begin{tabular}[c]{@{}c@{}}Permission Misuse: \\ Different Operation\end{tabular}           & 0.993              & 1               & 0.996             & 0.001        & 0            \\ \cline{2-7} 
   & Data Leakage     & 0.963              & 1               & 0.981             & 0.003        & 0    \\ \cline{2-7} 
  & \begin{tabular}[c]{@{}c@{}}Denial-of-Wallet:\\  Repeated Operation\end{tabular}    & 1     & 1               & 1                 & 0            & 0            \\ \cline{2-7} 
& \begin{tabular}[c]{@{}c@{}}Denial-of-Wallet: \\ Increased Function Duration\end{tabular}    & 0.976              & 1               & 0.988    & 0.002    & 0        \\ \hline
\multirow{5}{*}{\textbf{VOD}}    & \begin{tabular}[c]{@{}c@{}}Permission Misuse:\\ Different Order of Operations\end{tabular}  & 0.960   & 1     & 0.980              & 0.005        & 0      \\ \cline{2-7} 
& \begin{tabular}[c]{@{}c@{}}Permission Misuse:\\ Additional Operation\end{tabular}   & 0.893   & 1       & 0.944             & 0.006        & 0            \\ \cline{2-7} 
& Data Leakage     & 0.893   & 1    & 0.944    & 0.006        & 0         \\ \cline{2-7} 
 & \begin{tabular}[c]{@{}c@{}}Denial-of-Wallet:\\ Repeated Operation\end{tabular}     & 0.977              & 1               & 0.988             & 0.003        & 0   \\ \cline{2-7} 
 & \begin{tabular}[c]{@{}c@{}}Denial-of-Wallet:\\ Increased Function Duration\end{tabular}    & 0.969   & 1               & 0.984             & 0.004        & 0      \\ \hline
\end{tabular}%
}
\end{table}

The duration for each phase of the model, which includes training, testing, and retraining the model using our datasets, is presented in Table~\ref{tab:timing_table}.
This table also presents the average time required for training, testing, and retraining the model per individual sample, emphasizing the negligible time investment needed to apply the model on an extremely large dataset.

\begin{table}[h]
\caption{The number of instances for each model phase and the corresponding execution times
}
\renewcommand{\arraystretch}{1.15}
\label{tab:timing_table}
\resizebox{\columnwidth}{!}{%
\begin{tabular}{|c|c|c|c|}
\hline
\textbf{}           & \textbf{\begin{tabular}[c]{@{}c@{}}\# of Instances\end{tabular}} & \textbf{\begin{tabular}[c]{@{}c@{}}Time (sec)\end{tabular}} & \textbf{\begin{tabular}[c]{@{}c@{}}Time/Sample (sec)\end{tabular}} \\ \hline
\textbf{Training}   & 24,189                       & 994.959                                                           & 0.041                                                                          \\ \hline
\textbf{Testing}    & 23,147                       & 132.743                                                           & 0.006                                                                              \\ \hline
\textbf{Retraining} & 8,243                            & 325.731                                                                  & 0.039                                                                              \\ \hline
\end{tabular}%
}
\end{table}

\subsection{Updated Model Results} 

As mentioned earlier, our approach facilitates model updates by fine-tuning the model based on new data.
To achieve optimal outcomes during this updating process, based on empirical findings, we recommend constructing a training set that combines 90\% of new data with the remaining 10\% consisting of a random selection from the original training dataset.

To assess how well an updated model performs, we first trained and tested it using all the functions examined in the original experiment, except for the VOD-dynamo function, which was randomly excluded.
The results closely resembled those of the original model.

Following that, we constructed a new training dataset to update the model with.
The data from the 'new' VOD-dynamo function accounted for 90\% of the new training dataset, while the remaining 10\% was sourced from the initial training dataset, as detailed in Table~\ref{tab:data_table}.

The results of this experiment are presented in Table~\ref{tab:updated_model_results}.
In this case, the results remained consistent with the results obtained with the primary model.
This demonstrates the ability of the model to adapt to new information without compromising its performance.

\begin{table}[h]
\caption{Detection results for the initial training and subsequent model update}
\renewcommand{\arraystretch}{1.2}
\label{tab:updated_model_results}
\resizebox{\columnwidth}{!}{%
\begin{tabular}{|c|c|c|c|c|c|c|}
\hline
\textbf{Application}    & \textbf{Functions}   & \textbf{Precision} & \textbf{Recall} & \textbf{F1 Score} & \textbf{FPR} & \textbf{FNR} \\ \hline
\multirow{2}{*}{\textbf{\begin{tabular}[c]{@{}c@{}}Airline \\+ VOD\end{tabular}}} & \textbf{\begin{tabular}[c]{@{}c@{}}All functions \\ but one\end{tabular}} & 0.959              & 1               & 0.979             & 0.004        & 0            \\ \cline{2-7}                   & \textbf{\begin{tabular}[c]{@{}c@{}}All functions\end{tabular}}               & 0.967              & 1               & 0.983             & 0.003        & 0            \\ \hline
\end{tabular}%
}
\end{table}

\subsection{\label{sec:Discussion}Discussion} 

Some of the key strengths of our model are rooted in its generality (e.g., there is no need for a separate model per function or even per application) and ability to be fine-tuned based on new information.
In addition, our evaluation results suggest that the model is capable of scaling well (the performance remained similar despite expanding the model to more functions and adding another application).
Initially, our approach involved creating separate models for each function and every application. 
Each function was associated with its dedicated autoencoder, trained exclusively on the event sequences specific to that function's executions. 
Later, we expanded our approach by consolidating multiple functions under a single autoencoder. 
We started with functions from the same application and eventually extended this consolidation to encompass all event sequences, regardless of their function or application origins.

In order to maintain the quality of results achieved with individual autoencoders, we introduced two additional features to each event vector: the function name and the associated application name. 
However, our experimentation revealed that including the function name was unnecessary, as we were able to achieve nearly identical results by relying solely on the application name.
This enables the use of a single model, simplifying the application of our model while preserving its ability to effectively differentiate between benign and anomalous instances.

\section{Conclusion}
In this paper, we present a deep-learning threat detection model specifically designed for the identification of compromised serverless functions.
As the serverless application's activity logs are analyzed, our model scrutinizes them for any potential forensic evidence that might have been left behind by an attack that exploited serverless functions.
This is accomplished by using LSTM autoencoders to model the normal behavioral patterns of each serverless function. 
Consequently, our proposed method is threat-agnostic and designed to detect different types of threats, both known and unknown, across various threat categories.

Our method uses native hyper-logging capabilities provided by the serverless platform, and therefore it is self-contained and does not require any external information obtained through infrastructure modifications or third-party collaboration.
To evaluate our method's effectiveness, we created a serverless testbed consisting of two different applications that cover different typical serverless development paradigms.
We implemented several attack scenarios for each application in the testbed, demonstrating typical serverless threats against confidentiality, integrity, and availability.

Our evaluation demonstrates that it is possible to develop a threat detection model based solely on the information obtained from the cloud provider's \emph{native monitoring tools} that is capable of successfully detecting various attacks against serverless applications while maintaining a very low false alarm rate.
Furthermore, our approach has demonstrated its ability to be deployed across various applications and functions.
It exhibits adaptability through fine-tuning with new data, scalability, and utility for both online and offline detection scenarios.

\bibliographystyle{elsarticle-num}
\bibliography{bibli}

\end{document}